# High fidelity TiN processing modes for multi-gate Ge-based quantum devices

*Sinan Bugu[a*], Sheshank Biradar[a,b], Alan Blake[a], CheWee Liu[c], Maksym Myronov[d], Ray Duffy[a], Giorgos Fagas[a] and Nikolay Petkov[a,b*]*

[a] Tyndall National Institute, University College Cork, Lee Maltings, Cork, Ireland

[b] Physical Sciences Department, Munster Technological University (MTU), Rosa Avenue, Bishopstown, Cork, Ireland.

[c] Department of Electrical Engineering, National Taiwan University, Taiwan

[d] Department of Physics, University of Warwick, UK.

ABSTRACT Charge or spin-qubits can be realized by using gate-defined quantum dots (QDs) in semiconductors in a similar fashion to the processes used in CMOS for conventional field-effect transistors or more recent fin FET technology. However, to realize larger number of gate-defined qubits, multiples of gates with ultimately high resolution and fidelity is required. Electron beam lithography (EBL) offers flexible and tunable patterning of gate-defined spin-qubit devices for studying important quantum phenomena. While such devices are commonly realized by a positive resist process using metal lift-off, there are several clear limitations related to the resolution and the fidelity of patterning. Herein, we report a systematic study of an alternative TiN multi-gates definition approach based on the highest resolution hydrogen silsesquioxane (HSQ) EBL resist and all associated processing modes. The TiN gate arrays formed show excellent fidelity, dimensions down to 15 nm, various densities, and complexities. The processing modes developed were used to demonstrate applicability of this approach to forming multi-gate architectures for two types of spin-qubit devices prototypic to i) NW/fin-type FETs and ii) planar quantum well-type devices, both utilizing epi-grown Ge device layers on Si, where GeSn or Ge are the host materials for the QDs.

1. Introduction

There are several proposed platforms for realizing qubits, the basic units of quantum information processing, and to perform quantum computation [1]. While there has been great scientific progress and proof-of-concept demonstrations on all these platforms, to address the challenge of scalability it makes sense to use all the machinery of traditional semiconductor integrated circuits which also opens the possibility of integrating classical computing with quantum accelerators [2], [3]. To this end, recent developments in the field of quantum device engineering have shown that charge or spin-qubits can be realized in gate-defined quantum dots in semiconductors in a similar fashion to state-of-the art field-effect transistors (FETs) [4], [5], [6], [7]. In the last few years, Ge has progressed immensely from a conceptually new material for qubits to demonstrations of two-qubit logic and most recently first demonstration of a four-qubit quantum processor [8], [9]. Furthermore, the GeSn alloy is a material platform that carries the promise of highly desirable extreme mobility, low effective mass and added optical control for the qubits operation. The ultimate realization of quantum circuits will require patterning processes with high density gate structures at ultimately small widths and spacings between the neighboring structures to accommodate a large number of qubits [10].

Before such conceptually new devices are mass produced, CMOS-compliant patterning processes can be employed to demonstrate advantages promised. While extreme ultraviolet lithography (EUV) has been projected to reach ultimate resolution and fidelity for sub-10 nm CMOS technology modes [11], electron beam lithography (EBL) is currently used widely for device prototyping of Si and Ge spin-qubit devices [12]. Due to the easily tunable patterning, gate defined quantum dots (QDs) with controllable localization and density in Si, SiGe and Ge were realized to demonstrate quantum effects such as Coulomb blockade and spin exchange [9], [10]. Most of the demonstrated devices use metal gates including magnetic materials such as Cobalt [13], formed by lift-off process and defined in a positive polymethylmethacrylate (PMMA)-based EBL resist. While the PMMA EBL and lift-off processing modes are relatively well established there are several clear limitations related to the density/smallest dimensions obtainable, the line-edge roughness and the retention of metal between densely packed structures [12]. An alternative is the hydrogen silsesquioxane (HSQ), an inorganic resist, that is the resist of choice for high resolution definition of structures with sub-10 nm dimensions [14]. We have investigated the EBL HSQ process for epi-grown Ge-containing layers on Si substrates, demonstrating that moving from Si into Ge-containing substrates comes with its own challenges, limiting resolution and fidelity of the line structures obtained [15]. Titanium nitride (TiN) gate technology has already been introduced in CMOS fabrication capitalizing on the need for replacing ploy-Si gates with metal gates integrated with high-k gate oxides on Si but also for other channel materials [4], [5], [12]. Separately, TiN on its own can be classified as a quantum material as a result of its superconducting and plasmonic properties [16]. Extending this EBL-based technology to TiN for patterning gate-defined QDs and ultimately qubits is a natural development that capitalises upon ultimately small dimensions, high fidelity, and low line-edge roughness of the HSQ defined structures.

Herein we present a systematic study of the processing modes for realization of multi-gate TiN line structures, patterned by HSQ EBL and subsequent reactive ion etching (RIE). The TiN processing modes were specifically developed for Ge-containing substrates on Si, where GeSn or Ge are the host materials for the QDs, and for two types of quantum device architectures prototypic to i) NW/fin- FET devices and ii) planar quantum well (QW)-type devices, schematically depicted in Figure 1 a) and b), respectively. Each of the architectures presents its own separate challenges for the processing modes developed. For example, the NW devices are truly three dimensional (3D), using conformal deposition of the gate materials (oxide and TiN, metal) to form a π- or gate-all around gates. Notably, realizing multiple split gate architectures (Figure 1f) would not only double the total number of QDs but also provides unique quantization of the QDs at the corners of the NW channel [5]. For the QW-type devices utilizing planar heterostructure substrates (Figure 1b), the patterning challenges are in realizing large densities and varying complexities at ultimately small dimensions of the gates. Although, similar architectures have been demonstrated by using EBL in a positive resist and a subsequent metal lift-off, we demonstrate that the HSQ gate definition process brings all the advantages of extreme resolution and fidelity of patterning. Additionally, we estimated the resistivity of single line TiN structures with varying line widths and lengths to acquire initial electrical data of the TiN structures obtained.

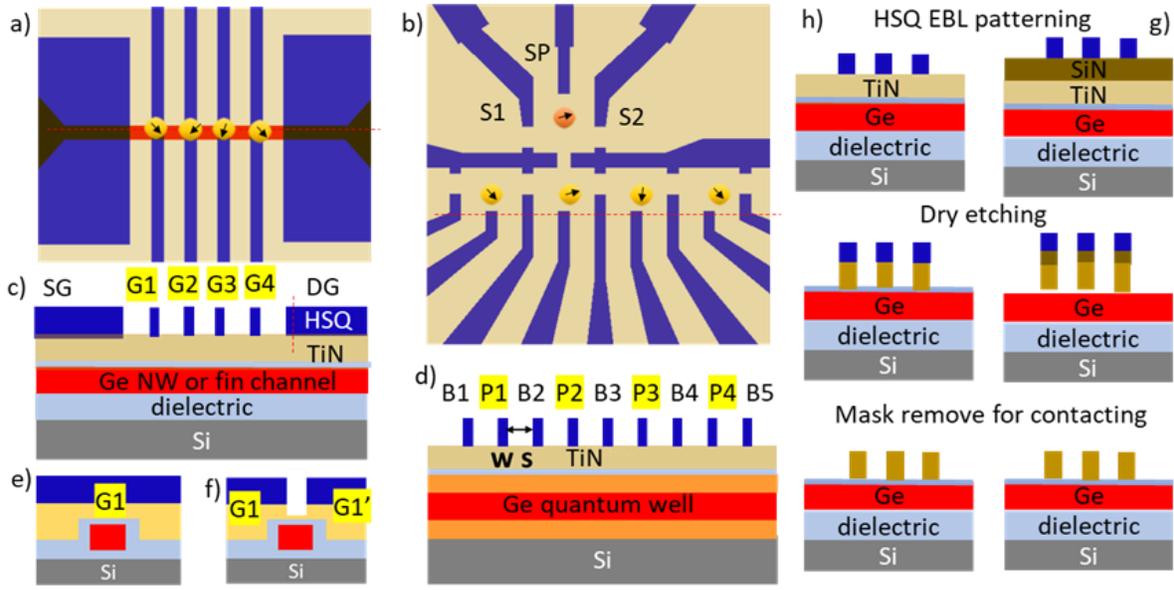

Figure 1. Schematics of the gate designs for gate-defined QDs (in yellow) using HSQ EBL processing for a) NW/fin FET devices and b) quantum well devices, and the corresponding cross-sectional representations of the layers c) source gate (SG), QD defining gates (G1-G4) and drain gate (DG) and in d) barrier gates (B1-B5) and plunger gates (P1 – P4). The representations in c) and d) are taken across the dotted lines. Shown also are the gates' width (W) and spacing (S) between the gates that are increased in the EBL design by 5 nm, starting at W = 15 nm and S = 2 × W. Additionally, the quantum well device in b) depicts a sensor dot defined by gates S1, S2 and sensor plunger gate (SP). For the NW or fin type devices in a) the gates can be defined as e) π-gates and f) split gates, latter doubles the total number of QDs. Summary of the total processing sequence for TiN gates definition using h) TMAH and g) salty NaOH HSQ developers; Ge is NW, fin or QW.

2. Materials and Methods

2.1 Fabrication modes

The processing steps for the realization of multi-gate TiN structures are schematically shown on Figure 1 g) and h) for TMAH and salty NaOH development processes, respectively. Note that the processing steps are different due to the need to introduce a sacrificial $SiN_x$ layer on top of the TiN for the salty NaOH HSQ process that, as we show below, is advantageous in achieving desired small dimensions and high fidelities of the structures. The use of the salty NaOH developer directly on the TiN surface was not successful due to HSQ lifting-off at any exposure dose used. We postulate that this is linked to the use of a the NaOH developer- highly concentrated strong base, that can oxidize and dissolve the surface titanium oxide and cause HSQ lift-off. Substrates used were pure Si, and Si substrates with epi-grown GeSn/Ge (550 nm thickness) and GeSi/GeQW/GeSi (about 3 µm thickness) layers. Previously, similar substrates were used to fabricate devices showing record mobilities with NW-FET [17] or 2D hole gas QW devices [18]. Before exposures substrates were covered by ALD with oxide (either 6 nm HfO or 20 nm $AlO_x$), and with 50 – 60 nm ALD TiN. The TiN films were deposited using a plasma-enhanced ALD process with a TDMA[Ti] precursor (heated to 90°C) in an ALD chamber (Cambridge Nanotech Fiji system) at constant pressure (300 mTorr) and temperature (250°C). Further details for the ALD process used can be found elsewhere [19]. For the substrates that were processed with the salty NaOH developer, a 35 nm PECVD $SiN_x$ was deposited on top of the TiN. The EBL exposures were done with two types of gate designs for gate-defined spin qubit devices as described in Figure 1a) and b. Figure SI1 from the Supporting Information shows example SEM images of the various gate designs investigated, that can be used for the realization of different number and topology of QDs part of the spin qubit devices. All images from Figure SI1 are for patterns developed on epi-grown Ge-containing layers. The patterns were exposed by using Elionix 100kV EBL system and 3 w/w % HSQ resist that was spin coated at 2000 rpm for 30 s and baked at 120 °C for 3 min to give 40 nm thick HSQ resist layer. The HSQ resist solution was freshly prepared using HSQ powder dissolved in dry methyl isobutyl ketone (MIBK) solvent, stored at 5°C and used within a week from its preparation for exposures. The exposures were done as dose tests starting at the minimal dose possible (determined by the resolution of the pattern generator and a beam step size of 0.5 nm) at a set current, i.e. when a 1nA e-beam current was used the minimal dose calculated was 4K $\mu C/cm^2$. The test structures had lines width (W) and spacing (S) that were varied starting at W = 15 nm and S = 2xW, increased by 5 nm up to W = 30 nm. The pattern transfer into the underlying substrates was done by $Cl_2$-based reactive ion etching (RIE) or consecutive Sulphur hexafluoride ($SF_6$) and $Cl_2$-based RIE steps. The HSQ resist removal to allow for contacting the TiN multi-gate structures was done by diluted buffered oxide etch (BOE) for 30 s, while the HSQ/$SiN_x$ was removed by combination of BOE and a $SF_6/O_2$ dry etch process. The final step was Ti/Au (100 nm) metal evaporation and lift-off in defined pads/lines to form metal contact pads for electrical testing.

2.2. Structural Inspection and Electrical testing

After every processing step, the substrates were imaged to obtain high resolution top-down SEM images on a Helios Nanolab dual beam SEM/FIB instrument using 5 kV acceleration voltage and through-lens detector. The instrument was also used to obtain cross-sectional SEM images of the patterned structures as well as energy dispersive X-ray (EDX) analysis with Oxford Instrument X-Max 100 detector with X-Max detector. Additionally, selected samples were sectioned by using the instrument to obtain thin lamella samples for cross-sectional STEM imaging. The site selective sample preparation followed well established protocols with final thinning of the foil at low Ga-ion beam energies. The lamella prepared were imaged using JEOL 2100 TEM and Titan STEM instruments, equipped with EDX detectors for elemental mapping.

To determine the resistivity of a material, a four-point contact measurement was conducted. Four Source Measure Units (SMUs) were used, with connections made at designated points on the sample (see the Supporting Information for further information on the electrical testing set-up used).

3.  Results and discussion

3.1 HSQ EBL patterning

The first step in the TiN gates definition process, schematically described in Figure 1 g) and h), is the HSQ EBL patterning. Briefly, lower dose e.g. 800 µC/cm$^2$ results in underexposed features falling over on their side walls, (marked in the Figure with white arrows), while at larger dose, the structures are not well resolved due to overexposure (indicated in the Figure with black arrows). These effects are seen across both FET and QW type architectures. Hence, the processing window for patterning gate structures with 15 nm width (W) and spacing between neighboring lines S = 2xW is very limited. In contrast, well developed structures with W = 30 nm are seen for both architectures, with the QW-type patterns requiring lower exposure dose due to the higher number of line structures and more complex architecture. In all cases, due to proximity effects (the presence of structures in proximity), the actual dose per line is enlarged that results in larger measured widths of the structures then the nominal and reduced S, correspondingly. To address the need for higher density (smaller W and S) gate structures, we have studied patterning using salty NaOH developer. The use of the salty NaOH developer required sacrificial SiNx, acting as additional hard mask to be introduced as explained in the Materials and Methods section. Figure 3 summarises the dose test results, showing that the smaller width (W = 15 nm) structures from both NW-type and QW-type designs, inaccessible by the TMAH developer, can be successfully patterned. There is a clear enhancement in the resolution and fidelity of the HSQ patterning, allowing high density gate arrays to be developed. This agrees well with the substantially improved resolution and fidelity of HSQ patterning ever since the salty NaOH developer was introduced [20], and further confirmation that its use does not preclude its application in CMOS manufacturing [21].

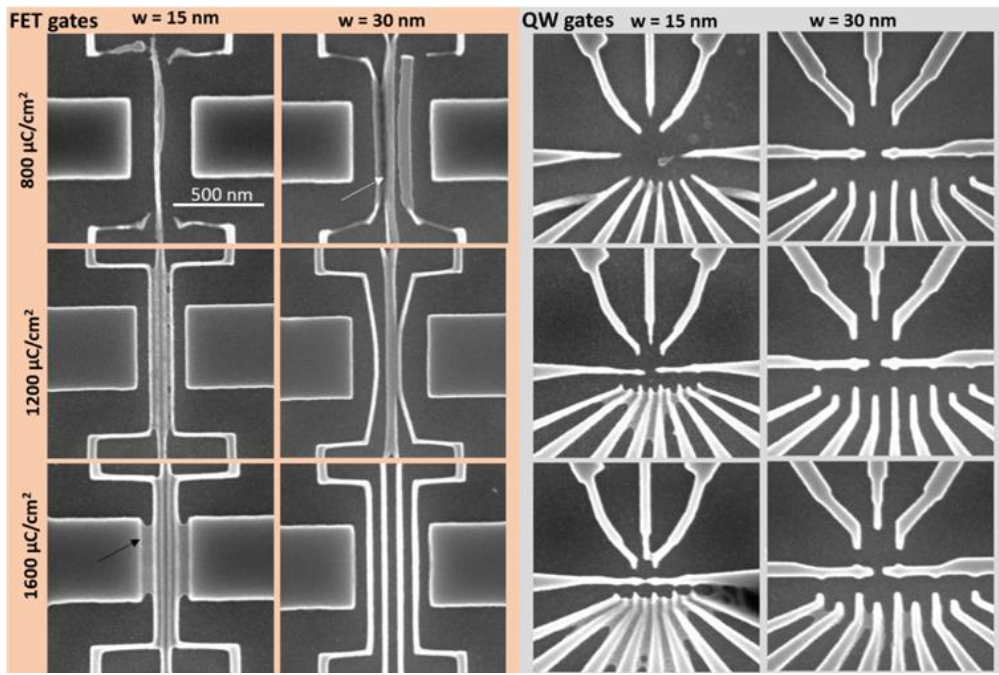

**Figure 2.** SEM images of HSQ dose tests at varying dose ($\mu C/cm^2$), developed by using TMAH developer for multi-gate FET (left, beige) and QW (right, grey) architectures. The designs shown are for widths (W) = 15 nm and 30 nm, and spacing (S = 2xW) exposed on Si, correspondingly.

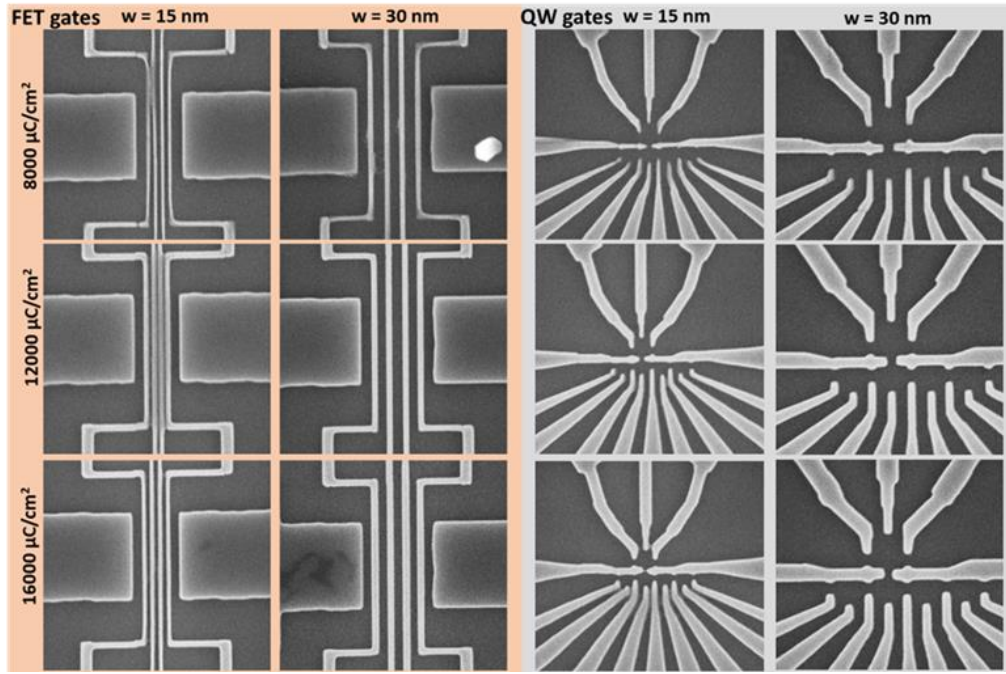

**Figure 3.** SEM images of HSQ dose tests at varying dose (µC/cm$^2$), developed by using salty NaOH developer for multi-gate FET (left, beige) and QW (right, grey) architectures. The designs shown are for widths (W) = 15 nm and 30 nm, and spacing (S = 2xW) exposed on Si, correspondingly.

Extending further the HSQ EBL process to epi-grown Ge-containing layers on Si (such are of interest to quantum devices as elaborated above), has some limitations as demonstrated by us before [15]. Briefly, we showed that at 100 kV incident e-beam, due to back-scattered electrons (BSEs) generation, the optimal range for high resolution/fidelity patterning is inversely proportional to the increase in the thickness (on the micro-scale) of the Ge-containing epi layers. Here we confirm that by introducing the gate stack (TiN/AlO$_x$) on the epi-Ge/Si substrates, the HSQ EBL exposures follow similar trend. The dose test exposures for the TiN/AlOx/Ge/GeSn/Ge/Si substrates with overall thickness of the Ge layers of 550 nm, show that the optimal dose range and fidelity of the structures is similar to the Si exposures (Figures 2 and 3) for both TMAH and salty NaOH developers, correspondingly (Supporting Information, Figures SI2 & SI3). This is not the case when the Ge-epi layers are with over a few micron thickness (data not shown) with largely overexposed features observed. Most importantly, Figures SI2 & SI3 once more demonstrate that the salty NaOH development is superior in achieving overall nominally small dimensions and high fidelity. Specifically, for the QW-well type devices where higher complexity and density of the gates are required, the HSQ salty NaOH development process offers unprecedented resolution and fidelity of the line structures in comparison to the other EBL patterning techniques. Moreover, to the best of our knowledge, Figure SI3 demonstrates the smallest W and S of array structures developed on GeSn/Ge containing substrates.

## 3.2 Pattern transfer in TiN by reactive ion-etching (RIE)

The next step in the TiN gates fabrication sequence is the pattern transfer into the underlying TiN layer. This is done by $Cl_2$-based RIE process for substrates where the HSQ structures are formed directly on the TiN, process sequence from Figure 1h. Figure 4a shows cross-sectional SEM image of a TiN array developed using 6 nm of hafnium oxide HfO as gate dielectric and an etch stop. Note that both the gate metal (TiN) and gate dielectric (HfO) are developed by ALD in the same chamber, and that the actual widths of the lines are larger due to proximity effect during patterning. Figures 4b and c depict TiN arrays etched by a consecutive application of $SF_6$-based ($SiN_x$ etch) and $Cl_2$-based (TiN) RIE recipes for substrates with sacrificial $SiN_x$ mask (patterning sequence from Figure 1g). The images show slightly over-etched $SiN_x$ structures, adopting typical mushroom-type morphologies, that are further transferred with good fidelity into the underlying TiN using 20 nm $AlO_x$ as an etch stop. With this process, denser arrays with good clearance between neighboring TiN lines and spacing of <30 nm can be achieved, (Figure 4b).

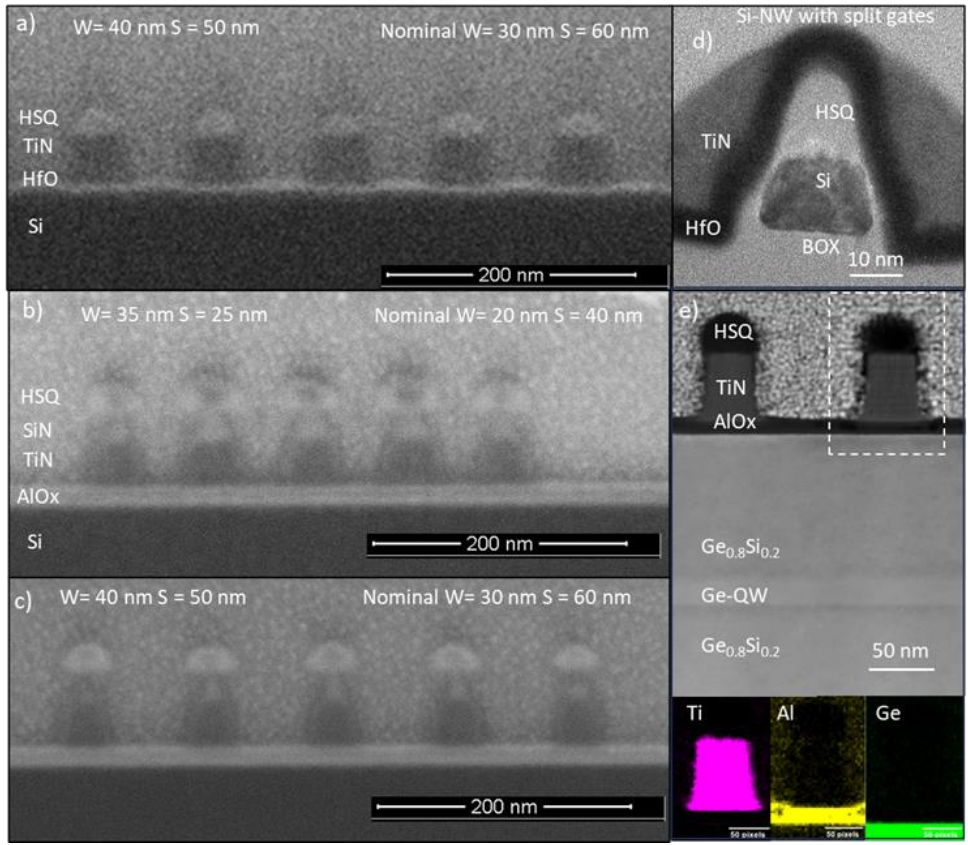

**Figure 4.** Selection of cross-sectional images demonstrating the high fidelity of the TiN RIE pattern transfer processes developed, a) SEM image of an array patterned using HSQ TMAH developer and etched using $Cl_2$-RIE with HfO as an etch stop, b) and c) SEM images of arrays patterned using salty NaOH developer and consecutive application of $SF_6$ ($SiN_x$ RIE) and $Cl_2$- RIE with $AlO_x$ as an etch stop. Insets depict measured and nominal dimensions of the arrays, note difference in the density of the arrays. d) STEM image of split-gate TiN/HfO FET device on SOI and e) STEM image of TiN/$AlO_x$ gates on $Ge_{0.8}Si_{0.2}$/GeQW/ $Ge_{0.8}Si_{0.2}$/Si substrate with corresponding EDX elemental mapping taken for one of the gates.

We have used the TiN etch processes discussed here to define gate lines for NW-type FET devices. Figure 5b shows a cross-sectional STEM image from a Si-NW FET device developed on silicon on insulator (SOI) substrate, where the TiN gates are formed as double split gates (schematics from Figure 1f). The image clearly shows all the nano-scale components of a split gate FET device with the Si channel, HfO gate dielectric conformally covering the channel at its sides and the TiN split gates that can be used to form QDs at the side walls or corners of the channel. Following similar process steps, we have also patterned fin-type architectures on GeSn/Ge substrates with single π- and doubled split-gates (Figure 5a and b). Additional, Figure 4e demonstrates the applicability of the HSQ TiN gates definition process to planar substrates such as $Ge_{0.8}Si_{0.2}$/GeQW/ $Ge_{0.8}Si_{0.2}$/Si that were used to previously to form high mobility two dimensional Ge-hole gas devices [18]. Specifically, the STEM image on Figure 4e shows two gates from an array of gates as shown on the top-down SEM image from Figure SI1d, whereby the Ti, Al, and Ge EDX spectral mapping validate the high fidelity of the etching process.

  The NW-type FET (Figure 4d) or the GeSn fin-type FET (Figure 5a and b) architectures require initial NW or fin definition, followed by conformal gate stack formation and TiN RIE etching to define the gates. Due to the non-planar topology provided by the side wall facets of the NW/fin channel, and the inherent high directionality of the RIE process, small TiN spacers (small side-wall residuals) can be formed, as evidenced by the cross-sectional SEM image on Figure 5c. The existence of such residuals will be detrimental to the performance of the multi-gate NW/fin FET devices as the gates will be electrically connected via the side-wall TiN spacers running parallel to the direction of the channel. To alleviate this, we developed an additional wet chemical TiN isotropic etch recipe that we refer to as TiN cleaning step. A similar approach has been used when investigating the wet etching of TiN in confined nano-spaces [22]. Briefly, the wet etch chemistry for this process is based on using conventional ammonia-peroxide mixture (APM) that is capable of nanometric removal of TiN by surface oxidation of the TiN to TiO2 and its dissolution under dilute ammonia. The completion of the etch was followed by EDX elemental line scans as depicted in Figure 5d-f. The TiN residuals were fully removed after 8 min of APM treatment, determined by the disappearance of the Ti signal in the EDX line scans, with the corresponding EDX spectrum plots shown on Figure 5f. Please note that the AlOx layer is not affected by the TiN cleaning step, as seen by the appearance of the Al signal in the EDX and additional cross-sectional images (data not shown).

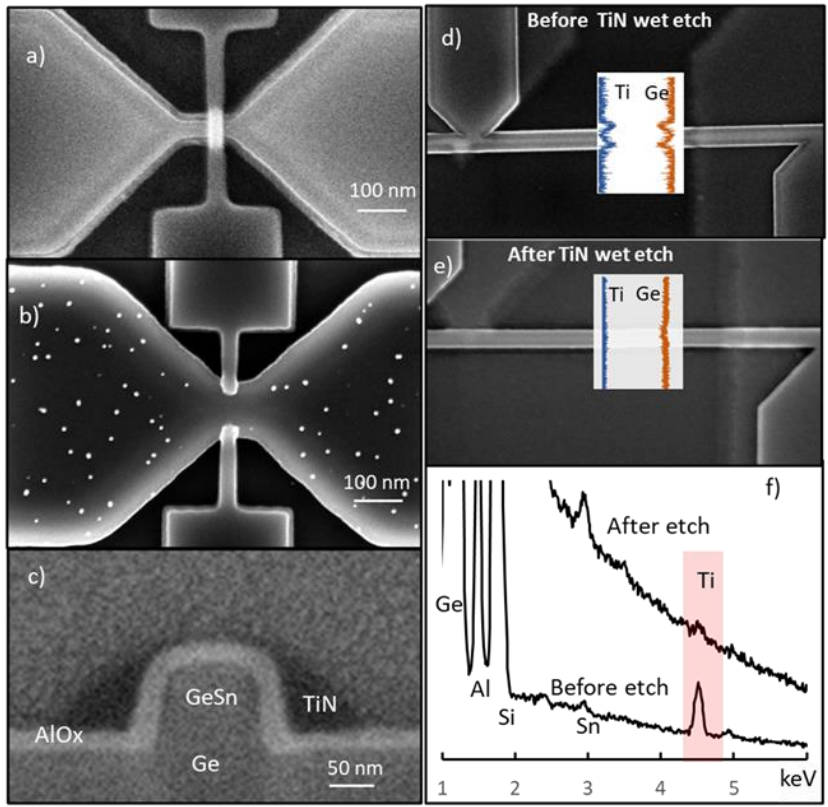

**Figure 5.** Top-down SEM images of the two main FET gate architectures developed by the TiN gate definition process a) single π-gate and b) double split-gate FET based on GeSn/Ge fin channel and conformal ALD $AlO_x$/TiN gate stack, c) corresponding cross-sectional SEM image in the area outside the gate depicting TiN residuals. Test GeSn/Ge fin structures after $AlO_x$/TiN ALD and TiN $Cl_2$-RIE d) before and e) after APM treatment (insets show corresponding EDX line scans

across the fins), and f) accompanying EDX sum spectra showing disappearance of the Ti-signal after APM treatment, note the appearance of the signals for Ge, Al, Si and Sn as expected.

3.3 Resist removal for metallization and electrical testing.

In order to develop metal contacts to the TiN gate structures, the EBL mask (HSQ or HSQ/SiN$_x$) removal needs to be implemented, as schematically shown on Figure 1h and g. That is done after optical lithography for contact pads definition followed by dilute BOE etch (HSQ removal) and it is further succeeded by a SF$_6$-based RIE when the process sequence started with HSQ/SiN$_x$ mask. Figure 6a shows a cross-sectional SEM image of a contact pad after BOE treatment and before metallization. It confirms that the HSQ is fully removed and that the overall TiN thickness was not substantially reduced after the treatment. AFM was used to evaluate and changes in the surface morphology in the TiN contact pad area after the HSQ removal step. The data shows that while the HSQ mask can be successfully removed, while the surface roughness of the underlying TiN is increased. Figure 6b depicts an example of a fully fabricated four terminals TiN resistor device having about 100 nm TiN contact line. Using such four terminal, single line resistor devices with varying width and length we obtained initial TiN resistivity data summarized in Figure 6c. The scatter plot shows that the resistivity values are in the range of about 150 – 700 μΩ-cm and agree well with the results of other published results for ALD deposited TiN films, including our own sheet resistance data [19], [23]. In all cases ohmic I/V curves were obtained (Figure SI 4b). For the range of diameters examined, the resistors data suggest no dependence on the TiN line width. Future electrical performance measurements including low temperature (down to mK regime) are planned to elucidate the performance of the TiN lines developed by the process modes described above as multi-gates for qubit devices.

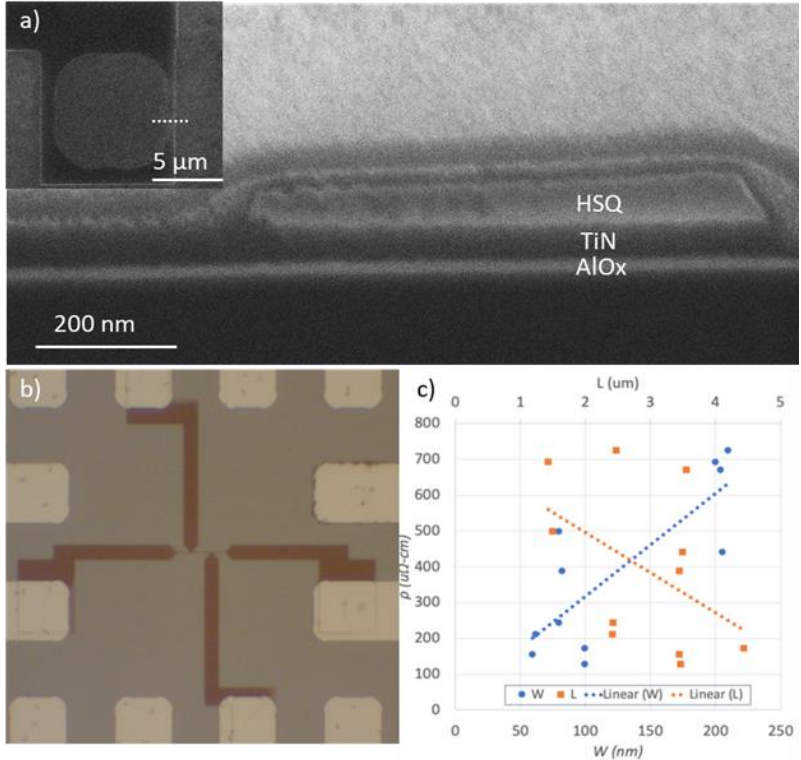

**Figure 6**. a) Cross-sectional SEM image taken in the contact pad area (inset in Figure 6a), b) optical microscopy image of a fully fabricated four terminals single line TiN resistor device fabricated following the processing sequence outlined in Figure 1g, d) scatter plot resistivity data for TiN single lines with varying width and length measured in a four-probe configuration.

4. Conclusions

In this report we formulate and systematically study the processing sequences for forming TiN multi-gate arrays with high resolution (down to 15 nm in width), density and fidelity. The processing modes based on EBL patterning were applied to define multi-gate architectures, that can be part of NW/fin FET and QW-type devices. Such devices can be used for realisation gate defined QDs, and further demonstrate spin-based qubits hosted in GeSn and Ge-QW. From a materials processing perspective, the gates definition processes described here are alternative to the commonly used metal lift-off gate definition, intended to allow for architectures with high complexity and applicability to various device concepts but also to provide high fidelity and resolution during pattering.

ASSOCIATED CONTENT

**Supporting Information**.

Additional top-down SEM images of the HSQ exposures at optimal dose for various multi-gate architectures and complexities, as well as at exposures at varying dose developed by using TMAH and salty NaOH developers on TiN/AlOx/GeSn/Ge on Si substrates. Four contact electrical testing set-up and I/V plots of single TiN resistors devices developed using the TiN processing modes studied.

AUTHOR INFORMATION

**Corresponding Author(s)**

\* Corresponding Author: Dr. Nikolay Petkov, email: Nikolay.Petkov@mtu.ie

\* Corresponding Author: Dr. Sinan Bugu, email: sinan.bugu@tyndall.ie**Author Contributions**

The manuscript was written through contributions of all authors. All authors have given approval to the final version of the manuscript.

ACKNOWLEDGMENT

This work was supported by the European Union's Horizon 2020 project ASCENT+ (GA No 871130) and Marie-Sklowdoska-Curie Fellowship (GA No 101066761), the Frontiers for17

SYNOPSIS TOC

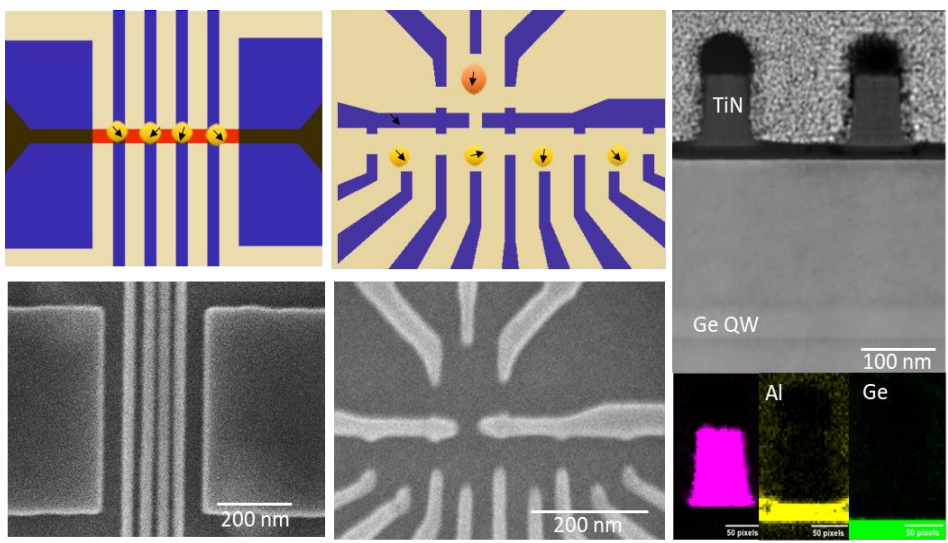

**Supporting information**

**High fidelity TiN processing modes for multi-gate Ge-based quantum devices**

Sinan Bugu[a*], Sheshank Biradar[b], Alan Blake[a], CheWee Liu[c], Maksym Myronov[d], Ray Duffy[a], Giorgos Fagas[a] and Nikolay Petkov[a,b*]

[a] Tyndall National Institute, University College Cork, Lee Maltings, Cork, Ireland

[b] Physical Sciences Department, Munster Technological University (MTU), Rosa Avenue, Bishopstown, Cork, Ireland.

[c] Department of Electrical Engineering, National Taiwan University, Taiwan

[d] Department of Physics, University of Warwick, UK.

Corresponding Author: Dr. Nikolay Petkov, email: Nikolay.Petkov@mtu.ie

Corresponding Author: Dr. Sinan Bugu, email: sinan.bugu@tyndall.ie

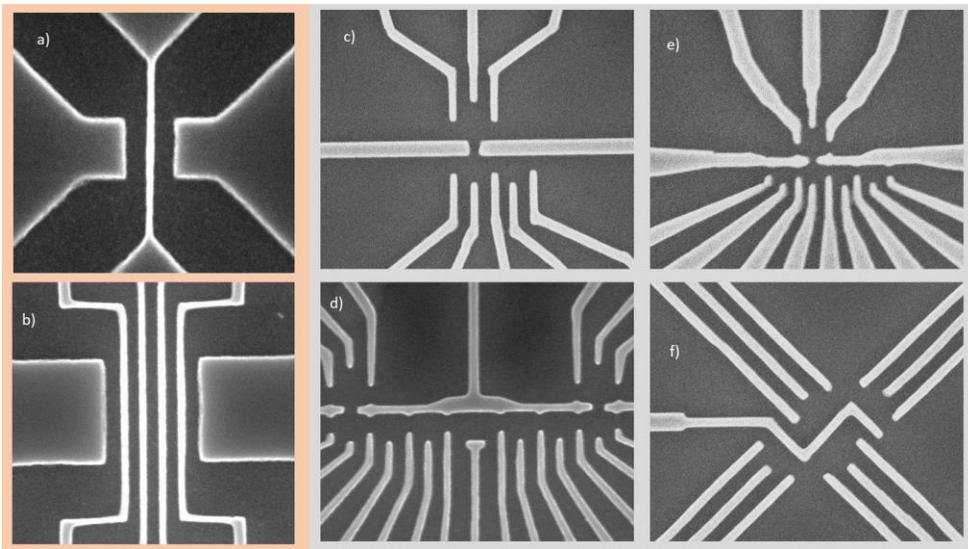

**Figure SI1**. SEM images of different type of gate designs with W = 30 nm and S = 60 nm, a) single and b) quadruple gates patterned with associated S/D gates on TiN/AlOx/GeSn/Ge on Si substrate for NW/fin FET devices, and c) – f) designs with varying number of gates and overall architecture patterned on TiN/AlOx/GeSi/GeQW/GeSi on Si substrate for QW-type devices. Note that all exposures are for epi-grown Ge device layers on Si substrates whereby the thickness of the Ge-containing layers are not exceeding 1 µm.

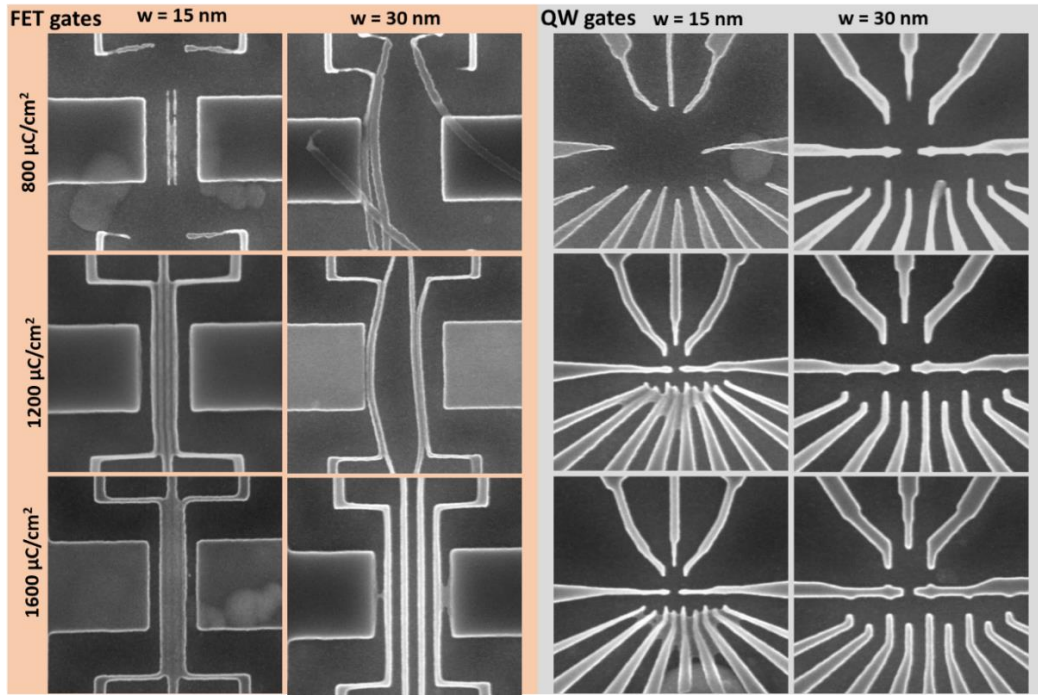

**Figure SI2**. SEM images of HSQ dose tests at varying dose (µC/cm2), developed by using TMAH developer for multi-gate FET (left, beige) and QW (right, grey) architectures. The designs shown are for widths (W) = 15 nm and 30 nm, and spacing (S = 2xW) exposed, TiN/AlOx/GeSn/Ge on Si substrates.

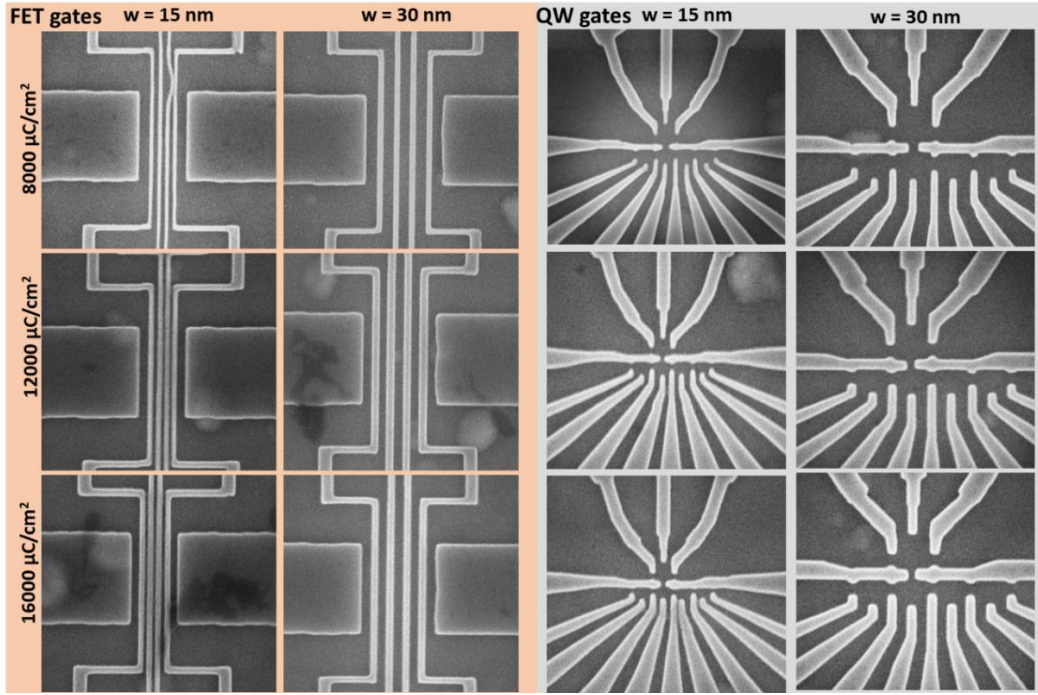

**Figure SI3.** SEM images of HSQ dose tests at varying dose (μC/cm$^2$), developed by using salty NaOH developer for multi-gate FET (left, beige) and QW (right, grey) architectures. The designs shown are for widths (W) = 15 nm and 30 nm, and spacing (S = 2xW) exposed, TiN/AlOx/GeSn/Ge on Si substrates.

*Electrical testing of resistor type devices*

The current source was programmed to sweep from -250 μA to +250 μA in 101 equal steps, ensuring both positive and negative currents were applied for comprehensive characterization. During each step of the current sweep, the voltage drop between two specific points, $V_2$ and $V_3$, was measured using the respective SMUs. Additionally, the voltage at point $V_4$ was monitored to ensure accurate reference measurements and stability of the test setup.

The collected data, plotting current ($I_{sweep}$) on the X-axis and the voltage drop ($V$) on the Y-axis, allowed for the determination of the material's resistance ($R$) by analyzing the slope of the I-V curve. The resistance obtained specifically from the area between $V_3$ and $V_4$ was used in subsequent calculations. This resistance, combined with the known dimensions of the sample, facilitated the accurate calculation of the TiN's resistivity. The resistivity ($\rho$) was calculated using the formula: $R = \frac{\rho L}{Wt}$ where $R$ is the measured resistance, $\rho$ is the resistivity, $L$ is the length between the voltage probes, $W$ is the width of the sample, and $t$ is the thickness of the sample. This method effectively minimized the influence of contact resistance and provided a precise measurement of the material's resistivity.

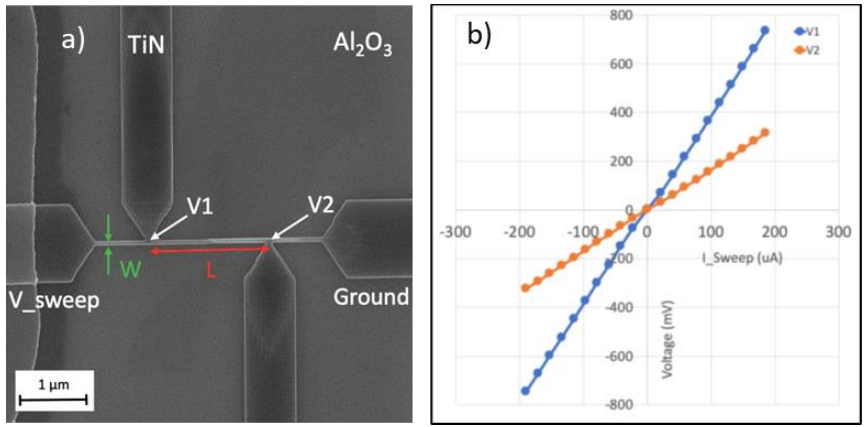

**Figure SI4.** a) SEM image of the four-point set up and b) corresponding I/V curves measured for a TiN wire with width (W = 80 nm).